\documentstyle[12pt]{article}
\newcommand{\Dslash}{\not \!\! D}

\begin{document}

%\begin{flushright}{May 22, 2007}
%\end{flushright}
%\vskip 0.5 truecm

\begin{center}
{\Large{\bf Non-hermitian radial momentum operator
 and path integrals
in polar coordinates}}
\end{center}
\vskip .5 truecm
\centerline{\bf Kazuo Fujikawa}

\vskip .4 truecm
\centerline {\it Institute of Quantum Science, College of 
Science and Technology}
\centerline {\it Nihon University, Chiyoda-ku, Tokyo 101-8308, 
Japan}
\vskip 0.5 truecm

%\makeatletter
%\@addtoreset{equation}{section}
%\def\theequation{\thesection.\arabic{equation}}
%\makeatother

\begin{abstract}
A salient feature of the Schr\"{o}dinger equation is that 
the classical radial momentum term $p_{r}^{2}$ in polar 
coordinates is replaced by the operator $\hat{P}^{\dagger}_{r}
\hat{P}_{r}$, where the operator $\hat{P}_{r}$ is not hermitian 
in general. This fact has important implications for the path
integral and semi-classical approximations. 
When one defines a formal hermitian radial momentum operator 
$\hat{p}_{r}=(1/2)\left((\frac{\hat{\vec{x}}}{r})
\hat{\vec{p}}+\hat{\vec{p}}(\frac{\hat{\vec{x}}}{r})\right)$, 
the relation $\hat{P}^{\dagger}_{r}
\hat{P}_{r}=\hat{p}_{r}^{2}+\hbar^{2}(d-1)(d-3)/(4r^{2})$ 
holds in $d$-dimensional space and this extra potential appears 
in the path integral formulated in polar coordinates.
The extra potential, which influences the classical solutions in 
the semi-classical treatment such as in the analysis of 
solitons and collective modes, vanishes for $d=3$ and attractive 
for $d=2$ and repulsive for all other cases $d\geq 4$. This 
extra term induced by the non-hermitian operator is a purely 
quantum effect, and it is somewhat analogous to the quantum 
anomaly in chiral gauge theory. 
\end{abstract}

%\large
\section{Introduction}

It is known that one needs to add an extra potential term to
the naive kinetic term in $d=2$ dimensions when one defines the 
path integral in polar coordinates~\cite{edwards}. This extra 
term has been later analyzed from a  point of view of path 
integrals in curved space~\cite{mclaughlin,mizrahi}. A connection
of this extra potential with the hermitian radial momentum 
operator has also been discussed in~\cite{arthurs}.
 This extra term is important in the quantum analyses of 
solitons~\cite{rajaraman} and collective modes~\cite{gervais}. 
The importance of this extra term was recently 
re-emphasized by Jackiw~\cite{jackiw}. 

In the present note, we discuss this problem  from a 
general point of view of the treatment of non-hermitian 
radial momentum operators in arbitrary dimensions.
Our basic observation is that the classical Hamiltonian in 
polar coordinates 
\begin{eqnarray}
H_{cl}=\frac{1}{2m}[p_{r}^{2}+\frac{\vec{L}^{2}}{r^{2}}]+V(r)
\end{eqnarray}
is replaced by the quantized Hamiltonian 
\begin{eqnarray}
\hat{H}=\frac{1}{2m}[\hat{P}^{\dagger}_{r}\hat{P}_{r}
+\frac{\hat{\vec{L}}^{2}}{r^{2}}]+V(r)
\end{eqnarray}
where the operator $\hat{P}_{r}$ is not hermitian in general and
$\hat{\vec{L}}^{2}$ stands for the quadratic Casimir operator 
of the rotation group in d-dimensional space. 
It is then shown that
\begin{eqnarray}
\hat{P}^{\dagger}_{r}\hat{P}_{r}=\hat{p}_{r}^{2}
+\frac{\hbar^{2}(d-1)(d-3)}{4r^{2}}
\end{eqnarray}
when one defines the formal hermitian radial operator 
$\hat{p}_{r}=(1/2)\left((\frac{\hat{\vec{x}}}{r})
\hat{\vec{p}}+\hat{\vec{p}}(\frac{\hat{\vec{x}}}{r})\right)$
in general $d$-dimensional space.
We then present an explicit construction of the path 
integral in polar coordinates starting with the quantum 
evolution operator.
It is  shown that the use of the hermitian or non-hermitian 
radial operator does not matter in the time slicing of the 
quantum evolution operator. But the formal hermitian operator 
gives a natural definition of the ``radial plane wave'', and 
thus it is essential to write the path integral in the 
conventional form. 

We also briefly mention that the appearance of an extra term induced by the non-hermitian operator and a technical aspect of the analysis are somewhat analogous to the quantum anomaly in chiral gauge theory.
    
\section{Hermitian radial momentum operator}

We start with the identity in the classical level in general $d$-dimensional space
\begin{eqnarray}
(\sum_{i}x_{i}p_{i})^{2}+\sum_{i\neq j}\frac{1}{2}(x_{i}p_{j}-
x_{j}p_{i})^{2}&=&\sum_{i,j}x^{2}_{i}p^{2}_{j}
=(\sum_{i}x^{2}_{i})(\sum_{j}p^{2}_{j})
\end{eqnarray}
Namely, we have the relation
\begin{eqnarray}
(\vec{p})^{2}=(\sum_{i}\frac{x_{i}}{r}p_{i})^{2}+\sum_{i\neq j}\frac{1}{2}\frac{1}{r^{2}}L^{2}_{i,j}
\end{eqnarray}
with
\begin{eqnarray}
&&r^{2}=\sum_{i}x^{2}_{i},\nonumber\\
&&L_{i,j}=x_{i}p_{j}-x_{j}p_{i},\ \ \ i\neq j.
\end{eqnarray}
We thus have the classical Hamiltonian
\begin{eqnarray}
H_{cl}&=&\frac{1}{2m}(\vec{p})^{2}+V(r)\nonumber\\
&=&\frac{1}{2m}p_{r}^{2}+\frac{1}{2m}\sum_{i\neq j}\frac{1}{2}\frac{1}{r^{2}}L^{2}_{i,j}+V(r)
\end{eqnarray}
with
\begin{eqnarray}
p_{r}=\sum_{i}\frac{x_{i}}{r}p_{i}.
\end{eqnarray}
One may define a general form of the quantized Hamiltonian by
\begin{eqnarray}
\hat{H}=\frac{1}{2m}\hat{P}_{r}^{\dagger}\hat{P}_{r}
+\frac{1}{2m}\sum_{i\neq j}\frac{1}{2}\frac{1}{r^{2}}
\hat{L}^{2}_{i,j}+V(r)
\end{eqnarray}
where $\hat{P}_{r}$ stands for the quantized version of 
$p_{r}$ in (8) which contains an operator ordering ambiguity.
The quantized $\hat{P}_{r}$ is not hermitian in general, and 
we transcribe the radial kinetic term by a quantized version 
$\hat{P}_{r}^{\dagger}\hat{P}_{r}$.  
The second term on the right-hand side of (9) does not contain 
any operator ordering problem since $\hat{L}_{i,j}$ itself has no
ordering ambiguity and 
\begin{eqnarray}
{[}\hat{L}_{i,j},r]=0
\end{eqnarray}
as $\hat{L}_{i,j}$ generates the rotation in the $i-j$ plane.

By noting the relation
\begin{eqnarray}
(\hat{\vec{p}})^{2}=\hat{P}_{r}^{\dagger}\hat{P}_{r}
+\sum_{i\neq j}\frac{1}{2}\frac{1}{r^{2}}
\hat{L}^{2}_{i,j}
\end{eqnarray}
the term $\hat{P}_{r}^{\dagger}\hat{P}_{r}$ needs to be positive
semi-definite, which is indeed the case by its construction
since $\hat{P}_{r}^{\dagger}\hat{P}_{r}\geq 0$ independently of 
the detailed definition of $\hat{P}_{r}$.
Secondly, $[(\hat{\vec{p}})^{2},c]=0$ for any constant $c$ and thus
\begin{eqnarray}
[\hat{P}_{r}^{\dagger}\hat{P}_{r},c]=0
\end{eqnarray}
should hold. This suggests that
\begin{eqnarray}
\hat{P}_{r}=\sum_{i}\frac{\hat{x}_{i}}{r}\hat{p}_{i}=
\frac{\hbar}{i}\frac{\partial}{\partial r}
\end{eqnarray}
and thus 
\begin{eqnarray}
\hat{P}_{r}^{\dagger}=\sum_{i}\hat{p}_{i}\frac{\hat{x}_{i}}{r}
=\frac{\hbar}{i}(\frac{\partial}{\partial r}+\frac{d-1}{r}).
\end{eqnarray}
We note that 
\begin{eqnarray}
{[}\hat{P}_{r}, r]=[\hat{P}_{r}^{\dagger}, r]=\frac{\hbar}{i}
\end{eqnarray}
and the general definition of $\hat{P}_{r}$ satisfies the
canonical commutation relation.
 
The quantized Hamiltonian is thus fixed to be
\begin{eqnarray}
\hat{H}&=&\frac{1}{2m}(\frac{\hbar}{i})^{2}
(\frac{\partial}{\partial r}+\frac{d-1}{r})
\frac{\partial}{\partial r}
+\frac{1}{2m}\sum_{i\neq j}\frac{1}{2}\frac{1}{r^{2}}
\hat{L}^{2}_{i,j}+V(r)\nonumber\\
&=&\frac{1}{2m}(\frac{\hbar}{i})^{2}\frac{1}{r^{(d-1)}}
\frac{\partial}{\partial r}r^{(d-1)}\frac{\partial}{\partial r}
+\frac{1}{2m}\sum_{i\neq j}\frac{1}{2}\frac{1}{r^{2}}
\hat{L}^{2}_{i,j}+V(r),
\end{eqnarray}
the radial part of which agrees with the radial part of the 
Laplacian in polar coordinates in general d-dimensional space.

 It is 
shown later that a formal hermitian radial momentum operator 
$\hat{p}_{r}$, 
which defines the ``radial plane wave'' naturally, is essential
to define the conventional form of the path integral for the 
radial component starting 
with the quantum evolution operator.
One may define the formal hermitian operator by
\begin{eqnarray}
\hat{p}_{r}&=&\frac{1}{2}\sum_{i}\{(\frac{\hat{x}_{i}}{r})
\hat{p}_{i}+\hat{p}_{i}(\frac{\hat{x}_{i}}{r})\}\nonumber\\
&=&\frac{\hbar}{i}\frac{1}{r^{\frac{(d-1)}{2}}}
\frac{\partial}{\partial r}r^{\frac{(d-1)}{2}}\nonumber\\
&=&\hat{p}^{\dagger}_{r}
\end{eqnarray}
in d-dimensional space. This operator $\hat{p}_{r}
=(\hat{P}_{r}+\hat{P}_{r}^{\dagger})/2$ also satisfies the 
canonical commutation relation
\begin{eqnarray}
{[}\hat{p}_{r}, r]=\frac{\hbar}{i}.
\end{eqnarray}
By using the relation
\begin{eqnarray}
(\hat{p}_{r})^{2}&=&-\hbar^{2}
\frac{1}{r^{\frac{(d-1)}{2}}}
\frac{\partial^{2}}{\partial r^{2}}{r^{\frac{(d-1)}{2}}}
\nonumber\\
&=&-\hbar^{2}\frac{1}{r^{(d-1)}}
\frac{\partial}{\partial r}r^{(d-1)}\frac{\partial}{\partial r}
-\hbar^{2}\frac{(d-1)(d-3)}{4}\frac{1}{r^{2}}
\nonumber\\
&=&\hat{P}_{r}^{\dagger}\hat{P}_{r}
-\hbar^{2}\frac{(d-1)(d-3)}{4}\frac{1}{r^{2}},
\end{eqnarray}
the quantized Hamiltonian (16) is finally written as 
\begin{eqnarray}
\hat{H}=\frac{1}{2m}\hat{p}_{r}^{2}+\frac{1}{2m}
\frac{\hbar^{2}(d-1)(d-3)}{4r^{2}}
+\frac{1}{2m}\sum_{i\neq j}\frac{1}{2}\frac{1}{r^{2}}
\hat{L}^{2}_{i,j}+V(r)
\end{eqnarray}
in terms of the hermitian radial momentum operator $\hat{p}_{r}$.
We note that $r$ in coordinate space and $p_{r}$ in momentum
space are not quite symmetric;~ $r=|\vec{r}|\geq 0$ but 
$\infty > p_{r}=\sum_{i}(x_{i}/r)p_{i}>-\infty$ and $|p_{r}|\neq
|\vec{p}|$ in general.
  
To find the values of the quadratic Casimir operator of the 
rotation group $SO(d)$
\begin{eqnarray}
\hat{C}_{2}=\sum_{i\neq j}\frac{1}{2}\hat{L}^{2}_{i,j},
\end{eqnarray}
we recall that the basis set defined by the $l$-th order 
homogeneous terms 
of $x_{1},.., x_{d}$, namely, $x^{l_{1}}_{1}x^{l_{2}}_{2}...
x^{l_{d}}_{d}$ with $\sum_{k}l_{k}=l$,  span an invariant space 
under 
the action of $\hat{L}_{i,j}$ which keeps $r$ invariant. We 
thus consider $(a_{1}x_{1}+ ... + a_{d}x_{d})^{l}=r^{l}Y_{l}$ 
with complex numbers $a_{1} \sim a_{d}$ which satisfy the 
condition~\footnote{If $a_{1}^{2}+ ... + a_{d}^{2}=0$ is not 
satisfied, $Y_{l}$ is mixed with $Y_{l-2}$ under the action of 
$\hat{C}_{2}$.},
$a_{1}^{2}+ ... + a_{d}^{2}=0$. We then have
\begin{eqnarray}
-\hbar^{2}\Delta (a_{1}x_{1}+ ... + a_{d}x_{d})^{l}&=&
[-\hbar^{2}\frac{1}{r^{(d-1)}}
\frac{\partial}{\partial r}r^{(d-1)}
\frac{\partial}{\partial r}+
\hat{C}_{2}\frac{1}{r^{2}}]r^{l}
Y_{l}\nonumber\\
&=&0
\end{eqnarray}
and thus
\begin{eqnarray}
\hat{C}_{2}Y_{l}
&=&\hbar^{2}l(l+d-2)Y_{l}.
\end{eqnarray}
The radial part of the Hamiltonian is thus written as 
\begin{eqnarray}
\hat{H}_{l}=\frac{1}{2m}\hat{p}_{r}^{2}+\frac{1}{2m}
\frac{\hbar^{2}(d-1)(d-3)}{4r^{2}}
+\frac{1}{2m}\frac{\hbar^{2}l(l+d-2)}
{r^{2}} + V(r).
\end{eqnarray}
The Casimir operator $\hat{C}_{2}$ is explicitly written in 
terms of angular variables in the polar coordinates such as
\begin{eqnarray}
&&x_{1}=r\cos\theta_{1},\nonumber\\
&&x_{2}=r\sin\theta_{1}\cos\theta_{2},\nonumber\\
&&x_{3}=r\sin\theta_{1}\sin\theta_{2}\cos\theta_{3},\nonumber\\
&&........\nonumber\\
&&x_{d-1}=r\sin\theta_{1}\sin\theta_{2}....\sin\theta_{(d-2)}
\cos\phi,\nonumber\\
&&x_{d}=r\sin\theta_{1}\sin\theta_{2}....\sin\theta_{(d-2)}
\sin\phi
\end{eqnarray}
with
\begin{eqnarray}
&&0\leq \theta_{1}\leq\pi, \ \ 0\leq \theta_{2}\leq\pi, \ \
....,\ \ 0\leq \theta_{(d-2)}\leq\pi,\nonumber\\
&& 0\leq \phi\leq 2\pi.
\end{eqnarray}
 
From the final expression of the quantized Hamiltonian (20) with
the formal hermitian radial momentum operator, we recognize that 
$d=3$ is exceptional in that the extra
potential vanishes, and $d=2$ is special in that the extra
potential is attractive. For all other cases $d\geq 4$, the 
extra potential is repulsive. This feature will be important 
when one considers a classical solution in the path integral as 
a starting point of the semi-classical 
analysis~\cite{rajaraman,gervais}.  

\section{Path integrals} 

To analyze the path integral for the radial coordinate, we start
with the definition of the eigenstates for the formal hermitian radial momentum  $\hat{p}_{r}$ in (17) by 
\begin{eqnarray}
\langle r|p_{r}\rangle =\frac{1}{r^{(d-1)/2}}\frac{1}{\sqrt{R}}
e^{ip_{r}r/\hbar}
\end{eqnarray}
which satisfies
\begin{eqnarray}
\langle r|\hat{p}_{r}|p_{r}\rangle=
p_{r}\langle r|p_{r}\rangle
=\frac{\hbar}{i}\frac{1}{r^{\frac{(d-1)}{2}}}
\frac{\partial}{\partial r}r^{\frac{(d-1)}{2}}
\langle r|p_{r}\rangle.
\end{eqnarray}

 The boundary condition for 
$\langle r|p_{r}\rangle$ may be chosen to be ``periodic'' inside
 a ball with a radius $R$, $0\leq r\leq R$, in the sense that
\begin{eqnarray}
&&
e^{ip_{r}0/\hbar}=
e^{ip_{r}R/\hbar},\ \ \ p_{r}=\frac{2\pi\hbar n}{R}, \ n=0, 
\pm 1,\pm 2, ..,
 \nonumber\\
&&\int^{R}_{0}r^{(d-1)}dr\langle r|p^{\prime}_{r}\rangle^{\star}\langle r|p_{r}\rangle=\int^{R}_{0}dr (\frac{1}{\sqrt{R}}
e^{ip^{\prime}_{r}r/\hbar})^{\star}\frac{1}{\sqrt{R}}
e^{ip_{r}r/\hbar}=\delta_{p_{r},p_{r}^{\prime}}
\end{eqnarray}
and let $R\rightarrow \infty$ later. This boundary condition
 ensures the hermiticity of $\hat{p}_{r}$~\footnote{When one 
imposes the conditions $r^{(d-1)/2}\psi(r)=0$ at $r=0$ and 
$r=\infty$, one can ensure the hermiticity of $\hat{p}_{r}$ in 
the sense $(\psi,\hat{p}_{r}\psi)=(\hat{p}_{r}\psi,\psi)$. But 
the eigenstates of $\hat{p}_{r}$ do not satisfy the boundary 
conditions. See, for example, \cite{galindo}. We thus define the 
complete set by (27) and (29)
with a periodic boundary condition for $r^{(d-1)/2}\psi(r)$
in the interval $0\leq r\leq R$.
This complete set provides the "radial plane waves" to define the 
path integral (33), and it gives  radial path integrals as 
defined in~\cite{edwards,arthurs} after the integral over
momentum variables. We do not assign a physical 
significance to eigenvalues $p_{r}$, and we use the radial 
plane waves just to define the path integral. (The $\delta$-functional source at $r=0$~\cite{galindo} is balanced by a sink at $r=R$ in the comlete set (29).) Note also that the
 plane waves in  cartesian coordinates, which are essential 
to define the path integral in  cartesian coordinates, do 
not necessarily satisfy the boundary condition of the 
relevant Schr\"{o}dinger wave function. We expect
 that our formulation is valid at least for the semi-classical 
approximation, which is the main physical 
interest~\cite{rajaraman, gervais} of the polar coordinate path
integral.}. 
We also have 
\begin{eqnarray}
\int_{-\infty}^{\infty} 
\frac{Rdp_{r}}{2\pi\hbar}\langle r_{1}|p_{r}\rangle\langle 
p_{r}|r_{2}\rangle=\frac{1}{(r_{1}r_{2})^{(d-1)/2}}
\delta(r_{1}-r_{2}).
\end{eqnarray}
The completeness relations are 
then written as 
\begin{eqnarray}
&&\sum_{l}\int_{-\infty}^{\infty} 
\frac{Rdp_{r}}{2\pi\hbar}|p_{r},l\rangle\langle 
p_{r},l|=1, \nonumber\\
&&\int_{0}^{R} r^{(d-1)}drd\Omega|r,\Omega\rangle\langle r,
\Omega|=1, 
\end{eqnarray}
where the symbols $\Omega$ and $l$  collectively stand for 
all the angular variables and all the quantum numbers associated
 with angular freedom, respectively. Note that $r=|\vec{r}|$ but
$p_{r}\neq |\vec{p}|$ and in fact we have
$-\infty< p_{r}<\infty$.

The path integral formula is written for 
$t=2\Delta t$, for example, in 
the following way. We first define
\begin{eqnarray}
\hat{H}_{l}&=&\frac{1}{2m}\hat{p}_{r}^{2}+\frac{\hbar^{2}}{2m}\frac{(d-1)(d-3)}
{4r^{2}}+\frac{\hbar^{2}l(l+d-2)}{2m r^{2}}+V(r)\nonumber\\
&\equiv&\frac{1}{2m}\hat{p}_{r}^{2}+\tilde{V}_{l}(r)
\end{eqnarray}
and then the conventional procedure by using the completeness
relations (30) and (31) gives
\begin{eqnarray}
&&\langle r_{f},l|e^{-\frac{i}{\hbar}\hat{H}2\Delta t}
|r_{i},l\rangle\nonumber\\
&&=\langle r_{f}|e^{-\frac{i}{\hbar}\hat{H}_{l}2\Delta t}|r_{i}
\rangle\nonumber\\
&&=\int_{0}^{R} r_{1}^{(d-1)}dr_{1}\langle r_{f}|
e^{-\frac{i}{\hbar}\hat{H}_{l}\Delta t}|r_{1}\rangle
\langle r_{1}|e^{-\frac{i}{\hbar}\hat{H}_{l}\Delta t}|r_{i}
\rangle\nonumber\\
&&=\int_{0}^{R} r_{1}^{(d-1)}dr_{1}\int_{-\infty}^{\infty}
\frac{Rdp_{r2}}{2\pi\hbar}
\frac{Rdp_{r1}}{2\pi\hbar}\langle r_{f}|
e^{-\frac{i}{\hbar}\hat{H}_{l}\Delta t}|p_{r2}\rangle
\langle p_{r2}|r_{1}\rangle
\nonumber\\
&&\times\langle r_{1}|e^{-\frac{i}{\hbar}\hat{H}_{l}\Delta t}
|p_{r1}\rangle
\langle p_{r1}|r_{i}\rangle\nonumber\\
&&=\frac{1}{\sqrt{(r_{f}r_{i})^{(d-1)}}}\int_{0}^{R} 
dr_{1}\int_{-\infty}^{\infty}\frac{dp_{r2}}{2\pi\hbar}
\frac{dp_{r1}}{2\pi\hbar}\exp\{\frac{i}{\hbar}
[(r_{f}-r_{1})p_{r2}
\nonumber\\
&&+(r_{1}-r_{i})p_{r1}-(\frac{p_{r2}^{2}}{2m}
+\frac{p_{r1}^{2}}{2m}+\tilde{V}_{l}(r_{f})+\tilde{V}_{l}(r_{1}))
\Delta t]\}
\nonumber\\
&&=\frac{1}{\sqrt{(r_{f}r_{i})^{(d-1)}}}
(\sqrt{\frac{m}{2\pi\hbar i\Delta t}}~)^{2}\int_{0}^{R} 
dr_{1}\nonumber\\
&&\times\exp\{\frac{i}{\hbar}[\frac{m}{2\Delta t}((r_{f}-r_{1})^{2}+(r_{1}-r_{i})^{2})-(\tilde{V}_{l}(r_{f})+\tilde{V}_{l}(r_{1}))
\Delta t]\}
\nonumber\\
&&=\frac{1}{\sqrt{(r_{f}r_{i})^{(d-1)}}}\langle r_{f}|
e^{-\frac{i}{\hbar}\hat{\tilde{H}}_{l}2\Delta t}|r_{i}\rangle
\end{eqnarray}
with
\begin{eqnarray}
\hat{\tilde{H}}_{l}&=&\frac{1}{2m}\hat{p}^{2}+\tilde{V}_{l}(r)
\nonumber\\
&=&-\frac{\hbar^{2}}{2m}\frac{\partial^{2}}{\partial r^{2}}
+\tilde{V}_{l}(r)
\end{eqnarray}
which stands for the Hamiltonian in one-dimensional space defined
by the interval $0\leq r\leq R$ with an effective potential 
$\tilde{V}_{l}(r)$. In (33) we used the standard procedure
\begin{eqnarray}
\langle r_{1}|e^{-\frac{i}{\hbar}\hat{H}_{l}\Delta t}
|r_{i}\rangle&=&\int \frac{Rdp_{r1}}{2\pi\hbar}
\langle r_{1}|e^{-\frac{i}{\hbar}\hat{H}_{l}\Delta t}
|p_{r1}\rangle\langle p_{r1}|r_{i}\rangle\nonumber\\
&\simeq&\int \frac{Rdp_{r1}}{2\pi\hbar}
\langle r_{1}|1-\frac{i}{\hbar}\hat{H}_{l}(\hat{p}_{r},\hat{r})\Delta t
|p_{r1}\rangle\langle p_{r1}|r_{i}\rangle\nonumber\\
&=&\int \frac{Rdp_{r1}}{2\pi\hbar}
\langle r_{1}|1-\frac{i}{\hbar}H_{l}(p_{r1},r_{1})\Delta t
|p_{r1}\rangle\langle p_{r1}|r_{i}\rangle\nonumber\\
&\simeq&\int \frac{Rdp_{r1}}{2\pi\hbar}\langle r_{1}|p_{r1}\rangle\langle p_{r1}|r_{i}\rangle
e^{-\frac{i}{\hbar}H_{l}(p_{r1},r_{1})\Delta t}
\end{eqnarray}
for an infinitesimal $\Delta t$. 

The expression for a general time interval is obtained by 
applying the composition law of the evolution operator to the 
expression (33), and one has
\begin{eqnarray}
\langle r_{f}|e^{-\frac{i}{\hbar}\hat{H}_{l}t}|r_{i}\rangle=
\frac{1}{\sqrt{(r_{f}r_{i})^{(d-1)}}}\langle r_{f}|e^{-\frac{i}{\hbar}\hat{\tilde{H}}_{l}t}|r_{i}\rangle
\end{eqnarray}
which relates the radial evolution operator in d-dimensional 
space on the left-hand side to the evolution operator in 
one-dimensional space on the right-hand side. Both expressions 
contain the effective potential with an extra term,
\begin{eqnarray}
\tilde{V}_{l}(r)=\frac{\hbar^{2}}{2m}\frac{(d-1)(d-3)}
{4r^{2}}+\frac{\hbar^{2}l(l+d-2)}{2m r^{2}}+V(r).
\end{eqnarray}

The path integral representation of the evolution operator for 
the radial freedom is thus formally written as 
\begin{eqnarray}
\langle r_{f},l|e^{-\frac{i}{\hbar}\hat{H}t}|r_{i},l\rangle&=&\langle r_{f}|e^{-\frac{i}{\hbar}\hat{H}_{l}t}|r_{i}\rangle\nonumber\\
&=&\frac{1}{\sqrt{(r_{f}r_{i})^{(d-1)}}}\int{\cal D}p_{r}{\cal D}r\exp\{\frac{i}{\hbar}\int_{0}^{t}dt
[p_{r}\dot{r}-H_{l}]\}\nonumber\\
&=&\frac{1}{\sqrt{(r_{f}r_{i})^{(d-1)}}}\int{\cal D}r\exp\{\frac{i}{\hbar}\int_{0}^{t}dt
[\frac{m}{2}\dot{r}^{2}-\frac{\hbar^{2}}{2m}\frac{(d-1)(d-3)}
{4r^{2}}\nonumber\\
&&\hspace{4cm} -\frac{\hbar^{2}l(l+d-2)}{2m r^{2}}-V(r)]\}
\end{eqnarray}
where the classical Hamiltonian $H_{l}$ is defined by 
\begin{eqnarray}
H_{l}=\frac{1}{2m}p_{r}^{2}+\frac{1}{2m}
\frac{\hbar^{2}(d-1)(d-3)}{4r^{2}}
+\frac{\hbar^{2}}{2m}\frac{l(l+d-2)}{r^{2}}+V(r)
\end{eqnarray}
and
\begin{eqnarray}
{\cal D}r\propto \prod_{t}dr(t)
\end{eqnarray}
without the naively expected weight factor 
$\prod_{t} r(t)^{(d-1)}$.
The classical solution in the semi-classical analysis is  
defined by the Lagrangian
\begin{eqnarray}
L_{l}=\frac{m}{2}\dot{r}^{2}-\frac{\hbar^{2}}{2m}
\frac{(d-1)(d-3)}
{4r^{2}} -\frac{\hbar^{2}l(l+d-2)}{2m r^{2}}-V(r)
\end{eqnarray}
and thus the classical solution is generally influenced by the 
extra term.

\section{Non-hermitian operator in path integrals}

To deal directly with the non-hermitian operator, we define 
the complete set of functions for the hermitian operator
$\hat{P}_{r}^{\dagger}\hat{P}_{r}$ defined by 
\begin{eqnarray}
\hat{P}_{r}^{\dagger}\hat{P}_{r}\langle r|P\rangle=
P^{2}\langle r|P\rangle.
\end{eqnarray}
We also use the notation $\varphi_{P}(r)=\langle r|P\rangle$
which satisfies 
\begin{eqnarray}
&&\int_{0}^{\infty} r^{(d-1)}dr \varphi^{\dagger}_{P}(r)
\varphi_{P^{\prime}}(r)=\frac{1}{P}\delta(P-P^{\prime}),
\nonumber\\
&&\int_{0}^{\infty} PdP \varphi^{\dagger}_{P}(r_{1})
\varphi_{P}(r_{2})=\frac{1}{(r_{1}r_{2})^{(d-1)/2}}
\delta(r_{1}-r_{2}).
\end{eqnarray}
We note that the radial coordinate $r$ and the radial momentum 
$P$ are more symmetric in the present definition, and the 
explicit
form of $\varphi_{P}(r)$ in $d=2$ is given by the Bessel function
$J_{0}(Pr)$  which satisfies
\begin{eqnarray}
\int_{0}^{\infty}PdP J_{0}(Pr_{1})J_{0}(Pr_{2})
=\frac{1}{\sqrt{r_{1}r_{2}}}\delta(r_{1}-r_{2}).
\end{eqnarray}

Then the evolution operator is written as 
\begin{eqnarray}
&&\langle r_{1},l|e^{-\frac{i}{\hbar}\hat{H}\Delta t}
|r_{i},l\rangle
\nonumber\\
&&=\langle r_{1}|e^{-\frac{i}{\hbar}\hat{H}_{l}\Delta t}
|r_{i}\rangle\nonumber\\
&&\simeq\int_{0}^{\infty}PdP
 \langle r_{1}|1-\frac{i}{\hbar}\hat{H}_{l}\Delta t
|P\rangle\langle P|r_{i}\rangle\nonumber\\
&&\simeq\int_{0}^{\infty}PdP \varphi_{P}(r_{1})
\varphi^{\dagger}_{P}(r_{i})\nonumber\\
&&\times \exp\{-\frac{i}{\hbar}[\frac{1}{2m}P^{2}
+\frac{\hbar^{2}}{2m}
\frac{l(l+d-2)}{r^{2}_{1}}+V(r_{1})]\Delta t\}
\end{eqnarray}
for an infinitesimal $\Delta t$. One can construct the evolution
operator for a finite time interval from (45) by using the 
composition law of the evolution operator, and
the expression on the right-hand side may be used for a 
numerical evaluation. But the expression on the right-hand side 
in (45) does not have an expression of the conventional path 
integral.
Since the expressions in (35) and (45) differ only in the choice 
of the complete states in the time slicing of the evolution
operator, these two expressions are equivalent to each other 
if one accepts the general operation
\begin{eqnarray}
\langle r_{1}|e^{-\frac{i}{\hbar}\hat{H}_{l}\Delta t}
|r_{i}\rangle
&\simeq&\sum_{n}
 \langle r_{1}|1-\frac{i}{\hbar}\hat{H}_{l}\Delta t
|n\rangle\langle n|r_{i}\rangle\nonumber\\
&=&\sum_{n}
 \langle r_{1}|1-\frac{i}{\hbar}H_{l}(n)\Delta t
|n\rangle\langle n|r_{i}\rangle\nonumber\\
&\simeq&\sum_{n}\langle r_{1}|n\rangle\langle n|r_{i}\rangle
e^{-\frac{i}{\hbar}H_{l}(n)\Delta t}
\end{eqnarray}
for an infinitesimal $\Delta t$.
But the analysis of  Edwards and Gulyaev \cite{edwards} suggests
 that the 
validity of this kind of operation is not always obvious.

It is thus instructive to derive the expression (35) from the 
last expression of (45) directly. By this way, we can also 
explain why no more extra potential is induced in (33).
For this purpose, we examine the object defined for Euclidean 
time 
\begin{eqnarray}
&&\int_{0}^{\infty}PdP \varphi_{P}(r_{1})
\varphi^{\dagger}_{P}(r_{i})
\exp\{-\frac{1}{\hbar}\frac{1}{2m}P^{2}\Delta\tau\}
\nonumber\\
&&=\int_{0}^{\infty}PdP
\exp\{-\frac{1}{\hbar}\frac{1}{2m}P^{2}\Delta\tau\}
 \varphi_{P}(r_{1})
\varphi^{\dagger}_{P}(r_{i})
\nonumber\\  
&&=\exp\{-\frac{1}{\hbar}
\frac{1}{2m}\hat{P}^{\dagger}_{r}\hat{P}_{r}\Delta\tau\} 
\int_{0}^{\infty}PdP\varphi_{P}(r_{1})
\varphi^{\dagger}_{P}(r_{i})\nonumber\\
&&=\exp\{-\frac{1}{\hbar}
\frac{1}{2m}\hat{P}^{\dagger}_{r}\hat{P}_{r}\Delta\tau\} 
\frac{1}{(r_{1}r_{i})^{(d-1)/2}}\delta(r_{1}-r_{i})\nonumber\\
&&=\exp\{-\frac{1}{\hbar}
\frac{1}{2m}\hat{P}^{\dagger}_{r}\hat{P}_{r}\Delta\tau\} 
\int_{-\infty}^{\infty}\frac{Rdp_{r}}{2\pi\hbar}
\langle r_{1}|p_{r}\rangle\langle p_{r}|r_{i}\rangle
\nonumber\\
&&=\exp\{-\frac{1}{\hbar}
\frac{1}{2m}[\hat{p}^{2}_{r}
+\frac{\hbar^{2}(d-1)(d-3)}{4r_{1}^{2}}
]\Delta\tau\} 
\int_{-\infty}^{\infty}\frac{Rdp_{r}}{2\pi\hbar}
\langle r_{1}|p_{r}\rangle\langle p_{r}|r_{i}\rangle
\nonumber\\
&&=\int_{-\infty}^{\infty}\frac{Rdp_{r}}{2\pi\hbar}
\langle p_{r}|r_{i}\rangle\exp\{-\frac{1}{\hbar}
\frac{1}{2m}[\hat{p}^{2}_{r}+\frac{\hbar^{2}(d-1)(d-3)}
{4r_{1}^{2}}
]\Delta\tau\} 
\langle r_{1}|p_{r}\rangle\nonumber\\
&&=\frac{1}{(r_{1}r_{i})^{(d-1)/2}}\int_{-\infty}^{\infty}
\frac{dp_{r}}{2\pi\hbar}\nonumber\\
&&\ \ \ \ \times e^{-ip_{r}r_{i}/\hbar}
\exp\{-\frac{1}{\hbar}
\frac{1}{2m}[\hat{p}^{2}+\frac{\hbar^{2}(d-1)(d-3)}
{4r_{1}^{2}}
]\Delta\tau\} 
e^{ip_{r}r_{1}/\hbar}
\end{eqnarray}
where we used (27) and the radial momentum operators act on the variable $r_{1}$. A new momentum operator
\begin{eqnarray}
\hat{p}=\frac{\hbar}{i}\frac{\partial}{\partial r_{1}}
\end{eqnarray}
was introduced in the last line of (47).

We thus examine
\begin{eqnarray}
&&\int_{-\infty}^{\infty}dp_{r}e^{-ip_{r}r_{i}/\hbar}
\exp\{-\frac{1}{\hbar}
\frac{1}{2m}\hat{p}^{2}\Delta -{\cal V}(r_{1})
\Delta\} 
e^{ip_{r}r_{1}/\hbar}
\end{eqnarray}
where we defined the simplifying notations
\begin{eqnarray}
&&{\cal V}(r_{1})=\frac{\hbar^{2}(d-1)(d-3)}{8\hbar m r_{1}^{2}},
\nonumber\\
&&\Delta=\Delta\tau.
\end{eqnarray}
The expression (49) is written as
\begin{eqnarray}
&&\int_{-\infty}^{\infty}dp_{r}e^{ip_{r}(r_{1}-r_{i})/\hbar}
\exp\{-\frac{1}{\hbar}
\frac{1}{2m}(p_{r}+\hat{p})^{2}\Delta-
{\cal V}(r_{1})\Delta\}\nonumber\\
&&=\int_{-\infty}^{\infty}\frac{dp_{r}}{\sqrt{\Delta}}
e^{ip_{r}\frac{(r_{1}-r_{i})}{\hbar\sqrt{\Delta}}
-\frac{p_{r}^{2}}
{2\hbar m}}
\exp\{-\frac{1}{\hbar m}\sqrt{\Delta}p_{r}\hat{p}
-\frac{1}{2\hbar m}\hat{p}^{2}\Delta-
{\cal V}(r_{1})\Delta\}\nonumber\\
&&=\int_{-\infty}^{\infty}\frac{dp_{r}}{\sqrt{\Delta}}
e^{ip_{r}\frac{(r_{1}-r_{i})}{\hbar\sqrt{\Delta}}
-\frac{p_{r}^{2}}
{2\hbar m}}
\exp\{-\frac{1}{\hbar m}\sqrt{\Delta}p_{r}\hat{p}
-{\cal V}(r_{1})\Delta\}
\end{eqnarray}
by moving the plane wave $e^{ip_{r}r_{1}/\hbar}$ through the 
operator and then re-scaling the integration variable as 
$p_{r}\rightarrow p_{r}/\sqrt{\Delta}$. We also used the fact 
that only the terms linear in $\Delta$ are important at the end.

We now expand the last exponential factor as 
\begin{eqnarray}
\exp\{-\frac{1}{\hbar m}\sqrt{\Delta}p_{r}\hat{p}
-{\cal V}(r_{1})\Delta\}=1+\sum_{n=1}^{\infty}\frac{(-1)^{n}}{n!}
(\frac{1}{\hbar m}\sqrt{\Delta}p_{r})^{n-1}\hat{p}^{n-1}
{\cal V}(r_{1})\Delta
\end{eqnarray}
which is replaced in the integrand of (51) after a shift of the 
integration variable $p_{r}\rightarrow p_{r}+i(r_{1}-r_{i})
\frac{m}{\sqrt{\Delta}}$ by 
\begin{eqnarray}
&&\int_{-\infty}^{\infty}\frac{dp_{r}}{\sqrt{\Delta}}
e^{-\frac{p_{r}^{2}}
{2\hbar m}}e^{-\frac{m(r_{1}-r_{i})^{2}}{2\hbar\Delta}}
\nonumber\\
&&\ \ \ \times[1+\sum_{n=1}^{\infty}\frac{(-1)^{n}}{n!}
(\frac{1}{\hbar m}\sqrt{\Delta}p_{r}+
\frac{i}{\hbar}(r_{1}-r_{i}))^{n-1}\hat{p}^{n-1}
{\cal V}(r_{1})\Delta]\nonumber\\
&&=\sqrt{\frac{2\pi\hbar m}{\Delta}}
e^{-\frac{m(r_{1}-r_{i})^{2}}{2\hbar\Delta}}
[1+\sum_{n=1}^{\infty}\frac{(-1)^{n}}{n!}
(r_{1}-r_{i})^{n-1}\frac{\partial^{n-1}}{\partial r^{n-1}}
{\cal V}(r_{1})\Delta]\nonumber\\
&&=\sqrt{\frac{2\pi\hbar m}{\Delta}}
e^{-\frac{m(r_{1}-r_{i})^{2}}{2\hbar\Delta}}
[1-\sum_{n=1}^{\infty}
(r_{1}-r_{i})^{n-1}\frac{\hbar^{2}(d-1)(d-3)}
{8\hbar m r_{1}^{n+1}}\Delta]
\end{eqnarray}
This last expression shows that 
$|r_{1}-r_{i}|\sim \sqrt{\Delta}$ and thus only the term with 
$n=1$ is important. We can thus replace (49) by
\begin{eqnarray}
\sqrt{\frac{2\pi\hbar m}{\Delta}}
e^{-\frac{m(r_{1}-r_{i})^{2}}{2\hbar\Delta}
-{\cal V}(r_{1})\Delta}=
\sqrt{\frac{2\pi\hbar m}{\Delta}}
e^{-\frac{m(r_{1}-r_{i})^{2}}{2\hbar\Delta}
-\frac{\hbar^{2}(d-1)(d-3)}{8\hbar m r_{1}^{2}}\Delta}
\end{eqnarray}
which establishes the equivalence of (45) with (35) and (33) 
after transforming back to the Minkowski metric,
$\Delta=\Delta\tau \rightarrow i\Delta t$.

This analysis shows that the use of the non-hermitian or 
hermitian radial momentum operator does not matter when 
defining the time slicing of the quantum evolution operator, 
but the formal hermitian radial operator is essential to write 
the path integral in the conventional form.

\section{Conventional path integrals in $d=2$}
We briefly discuss the conventional path integral for a free
particle in cartesian coordinates in $d=2$
\begin{eqnarray}
&&\langle \vec{x}_{f}|e^{-\frac{i}{\hbar}\hat{H}2\Delta t}
|\vec{x}_{i}\rangle\nonumber\\
&&=\int\frac{d^{2}p_{2}}{(2\pi\hbar)^{2}}\frac{d^{2}x_{1}
d^{2}p_{1}}{(2\pi\hbar)^{2}}\nonumber\\
&&\times\exp\{\frac{i}{\hbar}
[(\vec{x}_{f}-\vec{x}_{1})\vec{p}_{2}+(\vec{x}_{1}-\vec{x}_{i})
\vec{p}_{1}-\frac{1}{2m}((\vec{p}_{2})^{2}+(\vec{p}_{1})^{2})
\Delta t]\}.
\end{eqnarray}
The path integral for a general time interval is constructed 
from this expression by applying the composition law of the 
evolution operator, and it is known that this expression gives 
rise to an accurate evolution operator.

We thus examine the basic building block of the path integral
\begin{eqnarray}
\langle \vec{x}_{1}|e^{-\frac{i}{\hbar}\hat{H}\Delta t}
|\vec{x}_{i}\rangle=
\int\frac{d^{2}p_{1}}{(2\pi\hbar)^{2}}\exp\{\frac{i}{\hbar}
[(\vec{x}_{1}-\vec{x}_{i})\vec{p}_{1}-\frac{1}{2m}
(\vec{p}_{1})^{2}
\Delta t]\}
\end{eqnarray}
by writing the integration variables in polar coordinates.
We emphasize that this change of variables is for the {\em 
ordinary integral} and thus it should work without any 
complication. One may write (56) as 
\begin{eqnarray}
\int\frac{1}{|\vec{x}_{1}-\vec{x}_{i}|}\frac{dp_{r}dL}
{(2\pi\hbar)^{2}}
\exp\{\frac{i}{\hbar}[|\vec{x}_{1}-\vec{x}_{i}|
p_{r}-\frac{1}{2m}(p^{2}_{r}+\frac{L^{2}}{|\vec{x}_{1}-\vec{x}_{i}|^{2}})\Delta t]\}
\end{eqnarray}
by defining the variables 
\begin{eqnarray}
&&p_{r}=\frac{(\vec{x}_{1}-\vec{x}_{i})}
{|\vec{x}_{1}-\vec{x}_{i}|}\vec{p}_{1},\nonumber\\
&&L=\{(\vec{x}_{1}-\vec{x}_{i})\times\vec{p}_{1}\}_{3},
\end{eqnarray}
and one obtains the equality
\begin{eqnarray}
&&(\sqrt{\frac{m}{2i\pi\hbar\Delta t}}~)^{2}
\exp\{\frac{i}{\hbar}\frac{m}{2\Delta t}
|\vec{x}_{1}-\vec{x}_{i}|^{2}\}\nonumber\\
&&=(\sqrt{\frac{m}{2i\pi\hbar\Delta t}}~)^{2}
\exp\{\frac{i}{\hbar}\frac{m}{2\Delta t}[(r_{1}-r_{i})^{2}+
2r_{1}r_{i}(1-\cos\Delta\phi)]\}
\end{eqnarray}
where $\Delta\phi=\phi_{1}-\phi_{i}$. This result is the same 
as the direct evaluation of (56), and it is known that the 
fourth order term $\Delta\phi^{4}$ in
\begin{eqnarray}
2r_{1}r_{i}(1-\cos\Delta\phi)=r_{1}r_{i}[(\Delta\phi)^{2}
-\frac{1}{12}(\Delta\phi)^{4}]
\end{eqnarray}
is effectively replaced by the extra
 potential term $\hbar^{2}/(8mr^{2}_{1})$
when one integrates over $\Delta\phi$ as was 
shown by Edwards and Gulyaev~\cite{edwards}.

Alternatively, one may define
\begin{eqnarray}
\vec{x}_{1}\vec{p}_{1}&=&r_{1}p_{r},\nonumber\\ 
\vec{x}_{i}\vec{p}_{1}&=&|\vec{p}|r_{i}\cos(\phi+\Delta\phi)
\nonumber\\
&=&|\vec{p}|r_{i}\cos\phi\cos\Delta\phi-
|\vec{p}|r_{i}\sin\phi\sin\Delta\phi,
\end{eqnarray}
and thus 
\begin{eqnarray}
\vec{x}_{i}\vec{p}_{1}=r_{i}p_{r}\cos\Delta\phi
-\frac{L}{r_{1}}r_{i}\sin\Delta\phi
\end{eqnarray}
with $L=|\vec{p}|r_{1}\sin\phi$, 
and 
\begin{eqnarray}
(\vec{p})^{2}=p_{r}^{2}+\frac{L^{2}}{r^{2}_{1}}.
\end{eqnarray}
Note that $\phi=\phi_{p}-\phi_{1}$ and 
$\phi_{p}-\phi_{i}=\phi+\phi_{1}-\phi_{i}=\phi+\Delta\phi$.
Then
\begin{eqnarray}
&&\int\frac{d^{2}p_{1}}{(2\pi\hbar)^{2}}\exp\{\frac{i}{\hbar}[
(\vec{x}_{1}-\vec{x}_{i})\vec{p}_{1}-\frac{1}{2m}
(\vec{p}_{1})^{2}
\Delta t]\}\nonumber\\
&&=\int\frac{dL}{r_{1}(2\pi\hbar)^{2}}dp_{r}\exp\{\frac{i}{\hbar}
[(r_{1}-r_{i}\cos\Delta\phi)p_{r}
+\frac{L}{r_{1}}r_{i}\sin\Delta\phi\nonumber\\
&&\hspace{4cm}
-\frac{1}{2m}(p_{r}^{2}+\frac{L^{2}}{r^{2}_{1}})\Delta t]\}
\nonumber\\
&&\simeq\int\frac{dL}{\sqrt{r_{1}r_{i}}(2\pi\hbar)^{2}}dp_{r}
\exp\{\frac{i}{\hbar}
[(r_{1}-r_{i})p_{r}+L\Delta\phi\nonumber\\
&&\hspace{4cm}
-\frac{mr_{1}r_{i}}{24\Delta t}(\Delta\phi)^{4}
-\frac{1}{2m}(p_{r}^{2}+\frac{L^{2}}{r^{2}_{1}})\Delta t]\}
\end{eqnarray}
where the last line agrees with the second line when one 
integrates over $dL$ and $dp_{r}$, and the term with 
$(\Delta\phi)^{4}$ is effectively replaced by the extra 
potential term $\hbar^{2}/(8mr^{2}_{1})$ when one integrates 
over $\Delta\phi$ in the final path integral formula following 
the analysis of~\cite{edwards}. One can also 
write the integral over the angular 
momentum in (64) as 
\begin{eqnarray}
\frac{1}{\sqrt{\Delta t}}
\int_{-\infty}^{\infty} d\tilde{L}\exp\{\frac{i}{\hbar}
[\tilde{L}\Delta\phi/\sqrt{\Delta t}-\frac{1}{2m}
\frac{\tilde{L}^{2}}{r^{2}_{1}}]\}
\end{eqnarray}
by defining $\tilde{L}=\sqrt{\Delta t}L$.

The expression (64) may be compared with our result for $d=2$
, for example, in (35) 
\begin{eqnarray}
&&\langle r_{1},\phi_{1}|e^{-\frac{i}{\hbar}\frac{1}{2m}
(\hat{\vec{p}})^{2}\Delta t}|r_{i},\phi_{i}\rangle\nonumber\\
&&=\sum_{M}\int\frac{Rdp_{r}}{2\pi\hbar}
\langle r_{1},\phi_{1}|e^{-\frac{i}{\hbar}\frac{1}{2m}
(\hat{\vec{p}})^{2}\Delta t}|p_{r},M\rangle
\langle p_{r},M|r_{i},\phi_{i}\rangle\nonumber\\
&&=\sum_{M}\hbar\int\frac{1}{(2\pi\hbar)^{2}}
\frac{1}{\sqrt{r_{1}r_{i}}} dp_{r}
\exp\{\frac{i}{\hbar}[(r_{1}-r_{i})p_{r}
+\hbar M(\phi_{1}-\phi_{i})\nonumber\\
&&-\frac{1}{2m}(p_{r}^{2}-\frac{\hbar^{2}}{4r^{2}_{1}}
+\frac{\hbar^{2}M^{2}}{r^{2}_{1}})\Delta t]\}
\end{eqnarray} 
by noting
\begin{eqnarray}
&&\langle \phi|M\rangle=\frac{1}{\sqrt{2\pi}}e^{iM\phi},\ \ \ 
\int_{0}^{2\pi}d\phi\langle M^{\prime}|\phi
\rangle\langle\phi|M\rangle=\delta_{M,M^{\prime}}, \nonumber\\
&&\sum_{M=-\infty}^{\infty}\langle\phi|M\rangle
\langle M|\phi^{\prime}
\rangle=\delta(\phi-\phi^{\prime}).
\end{eqnarray}
The summation over $M$ in (66) is written as 
\begin{eqnarray}
&&\frac{1}{\sqrt{\Delta t}}\sum_{M}\hbar\sqrt{\Delta t}
\exp\{\frac{i}{\hbar}[\hbar\sqrt{\Delta t} M(\phi_{1}-\phi_{i})/
\sqrt{\Delta t}
-\frac{1}{2m}\frac{(\hbar\sqrt{\Delta t})^{2}M^{2}}
{r^{2}_{1}}]\}\nonumber\\
&&\rightarrow \frac{1}{\sqrt{\Delta t}}\int_{-\infty}^{\infty} 
d\tilde{L}
\exp\{\frac{i}{\hbar}[\tilde{L}(\phi_{1}-\phi_{i})/
\sqrt{\Delta t}
-\frac{1}{2m}\frac{\tilde{L}^{2}}{r^{2}_{1}}]\}
\end{eqnarray}
by defining $\tilde{L}=\hbar\sqrt{\Delta t} M$ for 
$\Delta t\rightarrow 0$, which agrees with (65).

From the comparison of (55) and (66) one concludes the following:
One can accurately translate the evolution operator into the 
path integral in cartesian coordinates as in (55), but one cannot
distinguish $\hat{p}_{r}^{2}$ and $\hat{P}^{\dagger}_{r}
\hat{P}_{r}$ in the classical expression $p_{r}^{2}$ and 
thus one cannot produce the extra potential from the 
{\em classical} expression
$(\vec{p})^{2}$. But the path integral (55) contains all the 
information and, in fact, the extra potential is reproduced 
by the integral over the angular variable $\phi$, as was 
shown by Edwards and Gulyaev~\cite{edwards}. On the other hand, 
the evolution 
operator is accurately translated into the conventional form of 
the path integral in polar
coordinates when one defines the formal hermitian radial momentum
operator as in (66). In this latter case, the extra potential 
is generated
by the difference of $\hat{p}_{r}^{2}$ and $\hat{P}^{\dagger}_{r}
\hat{P}_{r}$, and no more extra terms are generated from the 
angular variable as is shown in (68) and also in Section 4. 
See also the analysis by Arthurs~\cite{arthurs}.

\section{Discussion and conclusion}

We analyzed the problem of the extra potential,
which was originally discovered when writing the path integral 
in polar coordinates in $d=2$~\cite{edwards}, in a more
general setting. We emphasized 
that the extra potential term appearing in the path integral in 
polar coordinates in general d-dimensional space is purely a 
quantum effect associated with the 
non-hermitin radial momentum operator in the Sch\"{o}dinger 
problem, though the hermitian or non-hermitian radial momentum 
does not matter when defining the time slicing of the quantum 
evolution operator. We think that this phenomenon in the elementary 
Sch\"{o}dinger problem is important for the pedagogical purpose also.

In the following, we briefly mention an interesting analogy of 
this extra term with the quantum anomaly in chiral gauge theory.
In the original analysis of the chiral anomaly, the vector-like
gauge theory was analyzed and thus no explicit connection with
the non-hermitian operator appeared~\cite{bell, adler}.
In the path integral formulation of chiral 
anomalies~\cite{fujikawa}, in particular, in the analysis of 
chiral gauge theory, the non-hermitian Euclidean operator 
provides an interesting intuitive picture of the origin 
of quantum symmetry breaking. The general chiral gauge 
theory is defined by the Dirac action~\cite{bardeen} 
\begin{eqnarray}
\int d^{4}x{\cal L}=\int d^{4}x
\bar{\psi}(x)i\gamma^{\mu}(\partial_{\mu}-iV_{\mu}(x)
-iA_{\mu}(x)\gamma_{5})\psi(x)
\end{eqnarray}
in the background of the vector-like gauge field $V_{\mu}$ and
the axial-vector gauge field $A_{\mu}$; these fields may in 
general be 
non-Abelian Yang-Mills fields. In the Euclidean formulation 
of this problem, which defines the path integral more precisely,
one deals with the basic operator~\cite{fujikawa-suzuki}
\begin{eqnarray}
\Dslash=\gamma^{\mu}(\partial_{\mu}-iV_{\mu}(x)
-iA_{\mu}(x)\gamma_{5})
\end{eqnarray}
which is non-hermitian in the Euclidean sense
\begin{eqnarray}
\Dslash^{\dagger}=\gamma^{\mu}(\partial_{\mu}-iV_{\mu}(x)
+iA_{\mu}(x)\gamma_{5})\neq \Dslash.
\end{eqnarray}
One can maintain hermiticity if one replaces 
$A_{\mu}(x)\rightarrow iA_{\mu}(x)$, but then the axial gauge 
symmetry is spoiled.

One of the ways to analyze the anomaly in gauge symmetry is to 
analyze the responce of the Euclidean path integral
\begin{eqnarray}
Z(V_{\mu}, A_{\mu})=\int {\cal D}\bar{\psi}{\cal D}\psi
\exp\{\int d^{4}x {\cal L}\}
\end{eqnarray}
under the gauge transformation of $V_{\mu}$ and $A_{\mu}$.
Since the regularization is defined by the Euclidean operator 
${\Dslash}$, the quantum breaking of the axial gauge symmetry 
may be intuitively attributed to the non-hermitian property of the basic operator though this argument by itself does not explain the appearance of quantum anomaly.

Another way to analyze the chiral gauge anomaly is to examine the covariant form of the anomaly on the basis of the hermitian operator 
\begin{eqnarray}
&&\Dslash^{\dagger}\Dslash\varphi_{n}(x)
=\lambda^{2}_{n}\varphi_{n}(x),\ \ \ \ 
\Dslash\Dslash^{\dagger}\phi_{n}(x)
=\lambda^{2}_{n}\phi_{n}(x),
\end{eqnarray}
and the evaluation of the anomaly is reduced to the evaluation 
of~\cite{fujikawa-suzuki}
\begin{eqnarray}
Tr(\exp\{-\Dslash^{\dagger}\Dslash\Delta\tau\})=
\sum_{n}\varphi^{\dagger}_{n}(x)e^{-\lambda^{2}_{n}\Delta\tau}
\varphi_{n}(x)
\end{eqnarray}
which is analogous to (47). The extraction of the chiral anomaly
is achieved by using a plane wave in the evaluation of this 
trace or index. In this sense, the use of the eigenvectors for
the hermitian radial momentum operator (27) in the problem 
analyzed in the present paper, (47) and (53), corresponds to the use of the plane wave in the analysis of the chiral anomaly.
We also note that the actual evaluation of  chiral anomaly in field theory often uses the first quantized formulas in an essential way~\cite{peter}.

Although there are fundamental differences in 
the extra term we analyzed in the present paper and the chiral
gauge anomaly in field theory, we find it interesting that the non-hermitian operator plays an important role in 
defining quantum theory and a technical aspect of the analyses is similar in these two quite different phenomena.

\end{document}